\documentstyle[prc,aps,manuscript,eqsecnum]{revtex}

\begin{document}

\draft

\title{NUCLEAR SHAPE FLUCTUATIONS IN FERMI-LIQUID DROP MODEL}
\author{D. Kiderlen$^{1,2)}$, V. M. Kolomietz$^{1,3,4)}$ 
and S. Shlomo$^{4)}$}
\address{$^{1)}$Physik-Department TU M\"unchen, 85747 Garching,
Germany}
\address{$^{2)}$ National Superconducting Cyclotron Laboratory, 
Michigan State University, East Lansing, Michigan 48824}
\address{$^{3)}$Institute for Nuclear Research, 252022 Kiev, Ukraine}
\address{$^{4)}$Cyclotron Institute, Texas A\&M University, College
Station, TX 77843-3366, USA}

\maketitle

\begin{abstract}

Within the nuclear Fermi-liquid drop model, quantum and thermal 
fluctuations are considered by use of the Landau-Vlasov-Langevin
equation. The spectral correlation function of the nuclear surface 
fluctuations is evaluated in a simple model of an incompressible and 
irrotational Fermi liquid. The dependence of the spectral correlation 
function on the dynamical Fermi-surface distortion is established. 
The temperature at which the eigenvibrations become overdamped is 
calculated. It is shown that, for realistic values of the relaxation 
time parameter and in the high temperature regime, there is a particular 
eigenmode of the Fermi liquid drop where the restoring force is 
exclusively due to the dynamical Fermi-surface distortion.
\end{abstract}
 
\vskip 1cm


\section{Introduction}

The collective dynamics and the dissipative properties of the nuclear 
Fermi liquid depend in many aspects on the dynamical distortion of 
the Fermi surface in momentum space. As is well-known, taking into account
this distortion allows the description of a new class of
phenomena, the most famous of which are the giant multipole resonances.
Furthermore, the scattering of particles on the distorted Fermi surface
leads to relaxation of collective motion and gives  rise to nuclear 
viscosity \cite{ak}. This so-called collisional mechanism of relaxation is
influenced very strongly by the Pauli blocking in the scattering of nucleons 
close to the Fermi surface and depends on temperature and retardation 
effects \cite{lp,kmp}. We will consider below both of them, concentrating on
the fact that the retardation effects are especially important for
a proper description of the transition from the zero-sound regime to
the first-sound regime in the excited (heated) nucleus.

In the present paper, we are interested in the spectrum of fluctuations 
in shape variables. The precise form of such spectra can be expected to
depend on the parameters of the model, such as the collision time, and, 
especially, on the memory effects. Here, we want to study these dependencies
as one step to our ultimate goal of determining the model parameters from
a comparison with experimental data, as might be possible due to a
relation of the afore mentioned spectra to $\gamma-$spectra.

In what follows, we combine the thermal and quantum fluctuations by
means of the fluctuation-dissipation theorem. Such an
approach presents a convenient connection between different regimes
of collective motion such as the quantum zero-sound regime at 
zero temperature and the collisional first-sound regime
in a hot system. 

This paper is organized as follows. In section II we suggest a proof 
of the Langevin equation for nuclear shape variables, starting from 
the Landau-Vlasov kinetic equation. 
In the derivation, the main features of the dynamical distortion of the 
Fermi surface are taken into account. In section III, we discuss the
collision integral, for which a form is adopted that includes memory effects. 
The limiting cases of the zero-sound and the first-sound regimes are
discussed in section IV. Results of numerical calculations are presented 
in section V. We conclude and summarize in section VI. The Appendix provides 
a derivation of the relevant correlation functions for the fluctuating 
forces in the case of non-Markovian processes based on the (second) 
fluctuation-dissipation theorem. 

\section{Spectral correlation function for shape fluctuations in a 
Fermi-liquid drop}

To consider fluctuations of the collective variables in a Fermi-liquid 
drop we start from the following kinetic equation for the small deviation 
$\delta n$ of the distribution function  $n \equiv n({\bf r},{\bf p};t)$ 
from the one in equilibrium, $n_{eq}$,   
taking into account a random force \cite{ak}
\begin{equation}\label{2.1}
{\partial \over \partial t} \delta n + \hat {L} \delta n = 
I[\delta n] + y\, .
\end{equation}
On the left hand side of (\ref{2.1}), the operator $\hat L$ represents the 
drift terms including the selfconsistent mean field $U$,
$$
\hat {L} \delta n = {{\bf p} \over m} \cdot {\bf\nabla}_r \delta n -
{\bf\nabla}_r  U_{eq} \cdot {\bf\nabla}_p \delta n -
{\bf\nabla}_r  \delta U \cdot {\bf\nabla}_p n_{eq}\, .
$$
On the right hand side, $I[\delta n]$ is the linearized collision integral 
and $y \equiv y({\bf r},{\bf p};t)$ is a random variable representing the 
random force. As such, its ensemble average vanishes, 
$\langle y\rangle =0$, while its second moment can be related to properties
of the collision term, as shown in the Appendix.

To derive the equation of motion for the shape variables, 
we will follow the nuclear fluid dynamic approach of Ref. \cite{ns}, and
take into account the dynamic Fermi surface distortion up to multipolarity 
$l = 2$:
\begin{equation}\label{2.2}
\delta n = -{\Big({\partial n \over {\partial\epsilon}}\Big)_{eq}} 
\sum_{l,m_l}^{l=2} \delta n_{lm_l}({\bf r},t) Y_{lm_l}(\hat {\bf p}). 
\end{equation}
Here $\epsilon$ is the quasiparticle energy \cite{ak}. A generalization
of our approach to the case of an arbitrary multipolarity $l$
of the Fermi surface distortion can be done in a straightforward 
way, see Ref. \cite{kmp}.
Using Eqs. (\ref{2.1}) and (\ref{2.2}) we can derive a closed set of 
equations for the following moments of the distribution function, namely, 
local particle density $\rho$, velocity field $u_\nu$ and pressure tensor 
$P_{\nu\mu}$, in the form
\begin{equation}\label{2.3}
{\partial \over \partial t} \delta \rho = - \nabla_\nu (\rho_{eq}
u_\nu), 
\end{equation}
\begin{equation}\label{2.4}
m\rho_{eq} {\partial \over \partial t} u_\nu + \rho_{eq} \nabla_\nu 
\Big({\delta^2{\cal E} \over \delta \rho^2}\Big)_{eq} \delta \rho + 
\nabla_\mu P_{\nu \mu}^\prime = 0,
\end{equation}
\begin{equation}\label{2.5}
{\partial \over \partial t} P_{\nu \mu}^\prime + P_{eq}
(\nabla_\nu u_\mu + \nabla_\mu u_\nu - {2 \over 3}\delta_{\nu \mu}
\nabla_\alpha u_\alpha ) = I_{\nu \mu} + y_{\nu \mu},
\end{equation}
exploiting the conservation of particle number and momentum in 
collisions between the particles. Here $\cal E$ is the internal energy 
density, which is the sum of the kinetic energy density of the
Fermi motion and the potential energy density associated with the
nucleon-nucleon interaction. The equilibrium pressure of a Fermi
gas, $P_{eq}$, is given by 
\begin{equation}\label{2.6}
P_{eq} = {1 \over 3m} \int {d{\bf p} \over (2\pi \hbar)^3} p^2 n_{eq},
\end{equation}
$P_{\nu \mu}^\prime$ is the deviation of the pressure tensor from 
its isotropic part due to the Fermi surface distortion
\begin{equation}\label{2.7}
P_{\nu \mu}^\prime = -{1 \over m}\int {d{\bf p} \over (2\pi\hbar)^3} 
(p_\nu - mu_\nu)(p_\mu -mu_\mu)\Big( {\delta n \over \delta
\epsilon}\Big)_{eq} \sum_{m} \delta n_{2m}({\bf r},t) Y_{2m}(\hat {\bf p}) ,
\end{equation}
$I_{\nu\mu}$ is the second moment of the collision integral
\begin{equation}\label{2.8}
I_{\nu\mu} = {1 \over m} \int {d{\bf p} \over (2\pi\hbar)^3} p_\nu
p_\mu I[\delta n] 
\end{equation}
and $y_{\nu \mu}$ gives the contribution from the random force 
\begin{equation}\label{2.9}
y_{\nu\mu} = {1 \over m} \int {d{\bf p} \over (2\pi\hbar)^3} 
p_\nu p_\mu y .
\end{equation} 
Using the Fourier transformation for the pressure
\begin{equation}\label{2.10}
P_{\nu\mu}^\prime (t) = \int {d\omega \over 2\pi} e^{-i\omega t} 
P_{\nu\mu,\omega}^\prime  
\end{equation}
and similarly for the other time dependent variables we find the 
solution to Eq. (\ref{2.5}) as
\begin{equation}\label{2.11}
P_{\nu\mu,\omega}^\prime = {{i\omega \tau - (\omega \tau)^2} 
\over {1 + (\omega \tau)^2}} P_{eq} \Lambda_{\nu\mu,\omega} + 
{\tau \over {1 + (\omega \tau)^2}} (1 + i\omega \tau) 
y_{\nu\mu,\omega},
\end{equation}
where we used the symbol
\begin{equation}\label{2.12}
\Lambda_{\nu\mu,\omega} = \nabla_\nu \chi_{\mu,\omega} +
\nabla_\mu \chi_{\nu,\omega} - {2 \over 3}\delta_{\nu\mu}\nabla_
{\lambda}\chi_{\lambda,\omega} 
\end{equation}
for this combination of gradients of the Fourier transform 
$\chi_{\nu,\omega}$ of the displacement field. The time 
derivative of $\bbox{\chi}({\bf r}, t)$ is defined as the velocity 
field, hence
\begin{equation}\label{2.13}
u_{\nu,\omega} = -i\omega \chi_{\nu,\omega} .
\end{equation}
To obtain Eq. (\ref{2.11}) we have also used the fact that the tensor 
$I_{\nu\mu}$, Eq. (\ref{2.8}), can be reduced to 
\begin{equation}\label{2.14}
I_{\nu\mu,\omega} = -{1 \over \tau} P_{\nu\mu,\omega}^\prime ,
\end{equation}
due to our restriction to quadrupole deformation of the Fermi surface.
This is because the $l = 0$ and $1$ components of the expansion (\ref{2.2}) 
do not contribute to the collision integral, reflecting the conservation 
of particle number and momentum in a collision.  Note that the 
form (\ref{2.14}) is also correct for a Non-Markovian collision term, with
the collision time $\tau$ being dependent on the frequency $\omega$. 
(For convenience we will omit this frequency dependence of $\tau$ in
our notations.)

From Eqs. (\ref{2.3}), (\ref{2.4}) and (\ref{2.11}) we find the equation of 
motion for the displacement field $\chi_{\nu,\omega}$ in the form
\begin{equation}\label{2.16}
-\rho_{eq} \omega^2 \chi_{\nu,\omega} + \hat {\cal L} \chi_{\nu,\omega}
 = \nabla_{\mu}(\sigma_{\nu\mu,\omega} + s_{\nu\mu,\omega}),
\end{equation}
where the conservative terms are abbreviated by
\begin{equation}\label{2.17}
\hat {\cal L} \chi_{\nu,\omega} = -{1 \over m}\rho_{eq}\nabla_\nu 
\Big( {\delta^2{\cal E} \over \delta\rho^2}\Big)_{eq} \nabla_\mu
\rho_{eq}\chi_{\mu,\omega} - 
{\rm Im}\left({\omega\tau \over {1 -i\omega\tau}}\right)
\nabla_\mu {P_{eq} \over m} \Lambda_{\nu\mu,\omega} , 
\end{equation}
$\sigma_{\nu\mu}$ is the viscosity tensor
\begin{equation}\label{2.18}
\sigma_{\nu\mu,\omega} = -i(\omega /m) \eta 
(\omega) \Lambda_{\nu\mu,\omega}
\end{equation}
with the viscosity coefficient
\begin{equation}\label{2.19}
\eta (\omega) = {\rm Re}\left({\tau\over {1 -i\omega \tau}}\right) P_{eq}\, ,
\end{equation}
and $s_{\nu\mu,\omega}$ is the random pressure tensor
\begin{equation}\label{2.20}
s_{\nu\mu,\omega} = - {\tau (1 + i\omega \tau) \over 
m(1 + (\omega \tau)^2)} y_{\nu\mu,\omega} .
\end{equation}

The correlation properties of $s_{\nu\mu,\omega}$ can be obtained for 
the general case where we also take into account retardation and 
memory effects in the system, see the Appendix for details. Using the 
correlation properties of the random tensor $y_{\nu\mu,\omega}$, we 
find for the ensemble average of 
$$
{1 \over 2}[s_{\nu\mu,\omega}({\bf r}); 
s_{\nu^\prime \mu^\prime,\omega^\prime}({\bf r^\prime})]_+ =
{1 \over 2}\Big(s_{\nu\mu,\omega}({\bf r}) 
s_{\nu^\prime \mu^\prime,\omega^\prime}({\bf r^\prime}) +
s_{\nu^\prime \mu^\prime,\omega^\prime}({\bf r^\prime})
s_{\nu\mu,\omega}({\bf r})\Big)
$$
the result, see Eq. (A.17),
$$
{1 \over 2}\overline{[s_{\nu\mu,\omega}({\bf r}); 
s_{\nu^\prime \mu^\prime,\omega^\prime}({\bf r^\prime})]_+}
$$
\begin{eqnarray}\label{2.21}
= {4\pi \over m^2} E(\omega,T) \eta (\omega) \delta ({\bf r} - 
{\bf r^\prime}) \delta (\omega + \omega^\prime) [\delta_{\nu
\nu^\prime} \delta_{\mu \mu^\prime} + \delta_{\nu \mu^\prime} 
\delta_{\mu \nu^\prime} - {2 \over 3}\delta_{\nu \mu}
\delta_{\nu^\prime \mu^\prime}],
\end{eqnarray}
where
\begin{equation}\label{2.22}
E(\omega, T) = {\hbar\omega \over 2} \coth{\hbar\omega \over 2T}.
\end{equation}
We have pre<served the constant $\hbar$ in Eq. (\ref{2.22}) in order to
stress the fact that both quantum and thermal fluctuations are
involved in Eq. (\ref{2.21}) \cite{ll,fick}.

To apply (\ref{2.16}) to a finite system, we shall assume a sharp boundary 
of the Fermi liquid
\begin{equation}\label{2.24}
\rho = \rho_0 \Theta (R(t) - r).
\end{equation}
Below, we will also assume this simple form of the particle density
$\rho$, Eq. (\ref{2.24}), at non-zero nuclear temperatures $T \neq 0$.
In a hot nucleus, the particle density parameter $\rho_0$ in Eq. (\ref{2.24}) 
is temperature-dependent, $\rho_0 = \rho_0 (T)$. However, the form
(\ref{2.24}) ignores the existence of a nucleon vapour. 

For the description of small amplitude oscillations of a certain 
multipolarity $L$ of a liquid drop we specify the liquid surface as 
\begin{equation}\label{2.25}
r = R(t) = R_0 [1 + \sum_{M} \alpha_{LM}(t) Y_{LM}(\theta ,\phi)].
\end{equation}
We write the displacement field $\chi_\nu ({\bf r}, t)$ for an 
incompressible and irrotational flow, $\nabla_{\nu} \chi_{\nu} = 0$,
as \cite{ns} 
\begin{equation}\label{2.26}
\chi_\nu ({\bf r},t) = {\sum_{M} a_{LM,\nu}({\bf r}) \alpha_{LM}(t)} ,
\end{equation}
where
\begin{equation}\label{2.27}
a_{LM,\nu} ({\bf r}) = {1 \over LR_{0}^{L-2}} \nabla_\nu (r^L 
Y_{LM}(\theta ,\phi)) .
\end{equation}
Multiplying Eq. (\ref{2.16})  by  $ma^\ast_{LM,\nu}$, summing over $\nu$ 
and integrating over ${\bf r}$-space, we obtain the Langevin equation 
for the collective variables, 
\begin{equation}\label{2.28}
-\omega^2 B_L \alpha_{LM,\omega} + (C_{L}^{(LD)} + 
C_{L}^\prime)\alpha_{LM,\omega} - i\omega \gamma_L (\omega)
\alpha_{LM,\omega} = f_{LM,\omega}.
\end{equation}
The collective mass $B_L$ is found to be
\begin{equation}\label{2.29}
B_{L} = m \int d{\bf r} \rho_{eq} \sum_{\nu} \vert a_{LM,\nu}\vert^2 =
{3 \over 4\pi L} AmR_{0}^2 .
\end{equation}
The static stiffness coefficient $C_{L}^{(LD)}$ is derived from the 
first term on the right hand side of Eq. (\ref{2.17}) and is given by 
\cite{hm}
\begin{equation}\label{2.30}
C_L^{(LD)} = {1 \over 4\pi}(L-1)(L+2) b_S A^{2/3} - {5 \over 2\pi}
{{L-1} \over {2L+1}} b_C {Z^2 \over A^{1/3}},
\end{equation}
where $b_S$  and $b_C$ are the surface energy and Coulomb energy
coefficients appearing in the nuclear mass formula, respectively. 
This definition coincides with the one for the stiffness coefficient 
in the traditional liquid drop model for the nucleus. 
We point out, that the nucleon-nucleon interaction, manifested at
the starting equations (\ref{2.1}) and (\ref{2.2}), is presented in 
Eq. (\ref{2.28})  only implicitly through the phenomenological stiffness
coefficient $C_L^{(LD)}$. 

At finite frequencies, the distortion of the Fermi surface causes an 
additional  contribution, $C_{L}^\prime$ in Eq. (\ref{2.28}), to the 
stiffness coefficient. It is found as
\begin{equation}\label{2.31}
C_{L}^\prime \equiv C_{L}^\prime (\omega) 
 = {\rm Im}\left({\omega \tau\over {1 -i\omega\tau}}\right)
\int d{\bf r} P_{eq} \overline{\Lambda}_{\nu\mu}^{(LM)} 
\nabla_\mu a^\ast_{LM,\nu},
\end{equation}
where 
$$
\overline{\Lambda}_{\nu\mu}^{(LM)} = \nabla_\nu a_{LM,\mu} +
\nabla_\mu a_{LM,\nu} .
$$

Using $a_{LM,\nu}$ from Eq. (\ref{2.27}), we obtain for the integral in
Eq. (\ref{2.31}) 
\begin{equation}\label{2.32}
\int d{\bf r} {\rho_{eq}} \overline{\Lambda}_{\nu\mu}^{(LM)} \nabla_\mu 
a^\ast_{LM,\nu} = d_L \,\rho_0
\end{equation}
where the information about the multipolarity is in
$$
d_L = 2 {{(L-1)(2L+1)} \over L} R_{0}^3 .
$$
Thus we find
\begin{equation}\label{2.33}
C_{L}^\prime (\omega)
= d_L\,{\rm Im}\left({\omega\tau\over {1 -i\omega \tau}}\right)\, P_{eq} .
\end{equation}
The proportionality to $(\omega\tau)^2$, for small values of this
product, explains, why such a correction does not appear in the
hydrodynamic limit. 

For the friction coefficient $\gamma_L (\omega)$ in Eq. (\ref{2.28}) 
we obtain
\begin{equation}\label{2.35}
\gamma_L (\omega) = \eta (\omega) \int d{\bf r}\, 
\overline{\Lambda}_{\nu\mu}^{(LM)} \nabla_\mu a^\ast_{LM,\nu} =
d_L \, \eta (\omega) .
\end{equation}
Both, $C^\prime_L$ and $\gamma_L$ depend implicitly on the temperature
via the dependence of the collision time $\tau$ and of $R_0^3 \,P_{eq}$
on the $T$.

The random force $f_{LM,\omega}$ in Eq. (\ref{2.28}) is related to the
random pressure tensor $s_{\nu\mu,\omega}$ by 
\begin{equation}\label{2.36}
f_{LM,\omega} = -m\int d{\bf r} \,s_{\nu\mu,\omega} \nabla_\mu 
a^\ast_{LM,\nu} .
\end{equation}
Using Eqs. (\ref{2.21}) and (\ref{2.25}) we obtain the spectral correlation 
function $\overline{(f_{LM})_{\omega}^2}$ of the random force 
$f_{LM} (t)$: 
\begin{equation}\label{2.37}
\overline{(f_{LM})_{\omega}^2} = 2 \,E(\omega,T) \,\eta (\omega) 
\int d{\bf r} \,\overline{\Lambda}_{\nu\mu}^{(LM)} \nabla_\mu 
a^\ast_{LM,\nu} = 
2 \,E(\omega,T)\,\gamma_L (\omega) .
\end{equation}
The basic property of the random variable $y$, in Eq. (\ref{2.1}), 
$\overline{y} = \overline{y_{\nu\mu}}=0$ transfers to both, the
random pressure tensor, $\overline{s_{\nu\mu,\omega}}= 0$, and 
the random force, $\overline{f_{LM,\omega}} = 0$. 

Finally, using Eqs. (\ref{2.28}) and (\ref{2.37}) we can derive the spectral 
correlation function $\overline{(\alpha_L)_{\omega}^2}$ for shape 
fluctuations \cite{ll}:
\begin{equation}\label{2.43}
\overline{(\alpha_L)_{\omega}^2} =
{2\,E(\omega, T)\,\gamma_L(\omega) \over {B_{L}^2(\omega^2 - 
\omega_{L}^2(\omega))^2
+  \omega^2 \gamma_{L}^2(\omega)}} .
\end{equation}
The result Eq. (2.35) is of similar form as one would  obtain starting 
from hydrodynamics with the difference that here both the viscosity
and the eigenfrequency $\omega_L$ of the underdamped oscillator,
\begin{equation}\label{2.34}
\omega_L = \sqrt{(C_L^{(LD)} + C_L^\prime(\omega))/B_L}
\end{equation}
depend on the frequency themselves. This $\omega$ dependence is due to 
the deformation of the Fermi surface and the non-Markovian 
collision term.

\section{NON-MARKOVIAN COLLISION TERM}

We have already pointed out, that memory effects in the collision term
causes the collision time $\tau$, introduced in (\ref{2.14}), to depend
on frequency. In this section, we want to discuss this dependence in
more detail.

Below, we will take into account the main contributions to the 
relaxation time $\tau$, one arising from interparticle collisions (two-body 
dissipation with relaxation time $\tau_2$) and the other from collisions of 
nucleons with the moving nuclear surface (one-body dissipation with 
relaxation time $\tau_1$). Thus,
\begin{equation}\label{tau}
{1\over \tau} = {1\over \tau_1} + {1\over \tau_2}.
\end{equation}
The introduction of one-body dissipation reflects the peculiarities 
of our consideration: \\ 
1) The presence of the sharp edge in the finite nuclear Fermi-liquid 
drop leads to a renormalization of the collisional integral 
$I[\delta n]$ which is traditionally taken from  Fermi-liquid theory
of an infinite system; 2) The restriction to Fermi-surface distortions 
of multipolarity $l \leq 2$, see eq. (\ref{2.2}), does not allow us to 
take into consideration Landau damping. Thus the associated 
fragmentation width of collective excitations is missing. Introduction of 
one-body dissipation provides a phenomenological description of both 
contributions to the relaxation time.

The $\tau_1$ is related to the partial width $\Gamma_L^{(1)}$ 
describing the damping of the collective state due to one-body dissipation,
\begin{equation}\label{2.38}
\tau_1 = {2 \hbar \over \Gamma_L^{(1)}}.
\end{equation}
For the modified one-body dissipation \cite{ns,skn} the width 
$\Gamma_L^{(1)}$ is given by
\begin{equation}\label{2.39}
\Gamma_L^{(1)} = {1\over A}\,\pi \,(L-1)^2\,L\,\rho_0 \,
v_F \,\hbar \,\lambda^2 ,
\end{equation}
where $\lambda$ is a free parameter which we will take from a fit to the 
experimental data of the width of the multipole giant resonance (MGR).
For the case of finite temperatures, the Fermi velocity $v_F$ in
Eq. (\ref{2.39}) has to be replaced by the temperature dependent value,
see Ref. \cite{hpr}. The corresponding temperature dependent
correction to the one-body relaxation time $\tau_1$, Eq. (\ref{2.38}),
is small at $T << \epsilon_F$ and we will be neglected in this work. 

Let us now discuss the contribution of two particle collisions. One purpose 
of the present paper is to study modifications in the expressions for 
transport coefficients and fluctuations due to memory effects in the 
collision integral. These memory effects are realized in the dependence 
of $I[\delta n]$ not only on the distribution $\delta n$ at a given time 
but also on the value of $\delta n$ during earlier times. The weight with 
which the distribution $\delta n$ at previous times $t^\prime$ contributes 
to the value of $I[\delta n]$ at a given time $t$ is given by the kernel 
of the convolution integral representing $I[\delta n]$. In the present 
case, this kernel depends on the difference $t - t^\prime$ only, since 
we have linearized the collision integral with respect to equilibrium. 

The inverse collision time $1/\tau_2$, Eq. (\ref{2.14}), is the Fourier 
transform of the kernel mentioned and thus depends on the frequency. As a 
consequence of causality, $1/\tau_2$ is actually a complex function of 
$\omega$, whose 
real and imaginary part are related by the Kramers-Kronig relations.
Thus, inclusion of memory effects in the collision integral 
$I_{\nu \mu}$ in Eq. (\ref{2.5}) implies a consistent change of both 
dissipative and conservative forces in the equations of motion 
(\ref{2.16}) and (\ref{2.28}). 

A more detailed analysis of this question will be the subject of further 
investigation. Here, we only want to mention that we adopt phenomenological 
values for the parameters in our approach.  Therefore, for the frequency 
range in which we are interested in the present case, it seems possible to 
neglect the frequency dependent corrections to the conservative forces in 
Eq. (\ref{2.28}). 

This approximation corresponds to keeping the real part of $1/\tau_2$ only,
the inverse of which will, in the reminder, be called collisional relaxation 
time and be labeled by the same symbol, $\tau_2$. 
For frequencies small compared to the Fermi energy, one finds \cite{lp,kmp}:
\begin{equation}\label{2.15}
\tau_2 \equiv \tau_2(\omega,T) = {4\,\pi^2\,\beta \,
\hbar  \over {(\hbar \,\omega)^2 + \zeta T^2}}
\end{equation}
where $\beta$ and $\zeta$ are constant. For infinite nuclear matter 
$\beta$ can be related \cite{bp} to the  differential cross section for 
the scattering of two particles. The differences between the different
estimates of the parameter $\beta$ 
\cite{d,b,k,w,yan} are rather large:
$$
\beta =  2.4 \div 19.8 MeV.
$$
For $\zeta$ we will adopt the value of $\zeta = 4\,\pi^2$ \cite{ak,lp}.

\section{ZERO AND FIRST SOUND LIMIT}

Equation (\ref{2.43}) is valid for arbitrary collision times $\tau$ and 
thus describes both the zero and the first sound limit as well 
as the intermediate cases. From it one can obtain the leading order
terms in the different limits mentioned. 

{\it (1) First sound limit: $\omega\tau \rightarrow 0,\,\, T >> \hbar
\omega$} \\
The contribution from the dynamic distortion of the Fermi surface
can be neglected in this case and we have from Eq. (\ref{2.33}),
\begin{equation}\label{2.44}
C_{L}^\prime \approx 0 .   
\end{equation}
The eigenfrequencies $\omega_L$ of the shape oscillations are 
determined here by the usual liquid drop model as
\begin{equation}\label{2.45}
\omega_L^{(LD)} = \sqrt{C_{L}^{(LD)}/B_L} \,\, .
\end{equation}
In the high temperature regime, the Fermi liquid viscosity 
$\eta(\omega)$, Eq. (\ref{2.19}), approaches the classical expression 
\cite{ak}
\begin{equation}\label{2.46}
\eta = {1 \over 5} \rho_0 p_{F}^2 \tau (0) ,
\end{equation}
where $p_F$ is the Fermi momentum and $\tau (0) \equiv \tau (\omega =0)$. 
The spectral correlation function $\overline{(f_L)_{\omega}^2}$ 
of the random force can be found from Eqs. (\ref{2.37}), (\ref{2.22}) and 
(\ref{2.35}) 
\begin{equation}\label{2.47}
\overline{(f_L)_{\omega}^2} = 2 \gamma_L (0) T \, .
\end{equation}
This correlation function is independent of $\omega$, i.e., it 
corresponds to a white noise.

{\it (2) Zero sound regime: $\omega \tau \rightarrow \infty,\,\, 
T << \hbar \omega$} \\
The contribution to the stiffness coefficient from the dynamic 
distortion of the Fermi surface is now given by (see Eq. (\ref{2.33}))
\begin{equation}\label{2.48}
C_{L}^\prime (\omega) \approx \tilde{C}_L^\prime = d_L \, P_{eq}.
\end{equation}
This expression coincides with the analogous one from \cite{ns}.
We note that, in a cold Fermi liquid drop, $\tilde{C}_{L}^\prime$ 
provides the main contribution to the stiffness
coefficient. The viscosity coefficient $\eta(\omega)$, Eq. (\ref{2.19}),
can be approximated in this limit by
\begin{equation}\label{2.49}
\eta (\omega) = (P_{eq}/\kappa_0)\Big[1 + {1\over \omega^2}\,
\Big(\zeta \,T^2 + {\kappa_0 \over \tau_1}\Big)\Big],\,\,\,\,\,\,\,
\, \kappa_0 = 4\,\pi^2\,\beta \,\hbar \, {\rm c} .
\end{equation}
The random force spectral correlation function 
$\overline{(f_L)_{\omega}^2}$ is obtained from Eqs. (\ref{2.37}), 
(\ref{2.35}) and (\ref{2.22}) to be
\begin{equation}\label{2.50}
\overline{(f_L)_{\omega}^2} = \hbar \omega \tilde{\gamma}_L ,
\end{equation}
where
\begin{equation}\label{2.51}
\tilde{\gamma}_L = d_L \, P_{eq}/\kappa_0
\end{equation}
does not depend on $\omega$. The spectral correlation function (\ref{2.50}) 
now corresponds to a blue noise.

We recall that the quantum-mechanical zero-sound regime Eq. (\ref{2.49})
was obtained from the classical approach. It is due to the fact that 
the quantum fluctuations have been incorporated into the correlation 
function (\ref{2.21}) through the factor $E(\omega,T)$, Eq. (\ref{2.22}), 
see also Refs. \cite{ll,fick}.

\section{Results and Discussion}

For the numerical calculations in this work we adopt the value of
$r_0 = 1.12\,{\rm fm}$  and assume a temperature dependence of 
the surface and Coulomb parameters in the liquid drop stiffness 
coefficient $C_L^{(LD)}$ of Eq. (\ref{2.30}), namely 
\cite{gsb,rpl}
\begin{equation}\label{3.1}
b_S = 17.2\,\Big({{T_C^2 - T^2}\over {T_C^2 + T^2}}\Big)^{5/4}\, 
{\rm MeV} ,\,\,\,\,\,b_C = 0.7 (1 - x_C T^2) \,{\rm MeV},
\end{equation}
where the parameter $x_C$ was chosen as 
$x_C = 0.76 \cdot 10^{-3} \,{\rm MeV}^{-2}$ \cite{gsb} and 
$T_C = 18 \,{\rm MeV}$ is taken as the critical temperature 
for infinite nuclear Fermi-liquid \cite{rpl}.
Using Eq. (\ref{3.1}), one can find a critical temperature $T_{in}^{(LD)}$
where the liquid drop contribution $C^{(LD)}_L$ to the stiffness coefficient 
vanishes:
\begin{equation}\label{3.2}
C_L^{(LD)} \equiv C_L^{(LD)}(T)\Big\vert_{T=T_{in}^{(LD)}} = 0.
\end{equation}
For the parameters used in the present work one obtains 
$T^{(LD)}_{in}= 7.72\, {\rm MeV}$ for quadrupole deformation, $L=2$, in 
$^{208}Pb$. For temperatures $T>T^{(LD)}_{in}$ there exists always an 
eigenfrequency with a positive imaginary part giving rise to an 
exponentially growing deformation. 

This eigenfrequency, along with possible others, are solutions
of the following secular equation, see Eq. (\ref{2.28}),
\begin{equation}\label{3.3}
-\omega^2\,B_L + \Big(C_L^{(LD)} + C_L^\prime (\omega)\Big)
- i\,\omega\,\gamma_L(\omega) = 0 .
\end{equation}
The transport coefficients $C_L^\prime (\omega)$ and $\gamma_L(\omega)$
are $\omega$-dependent because of the memory effects. To solve Eq. 
(\ref{3.3}), both coefficients have to be defined in the complex 
$\omega$-plane through analytical continuation of the corresponding 
expressions (\ref{2.33}) and (\ref{2.35}). In Fig. 1 we show the 
real (${\rm Re}\,\omega$) and imaginary (${\rm Im}\,\omega$) parts 
of the eigenfrequencies, obtained by solving
Eq. (\ref{3.3}) as functions of the temperature for the nucleus with 
$A = 208$, using $\zeta= 4\pi^2$, Eq. (\ref{2.15}), and  three different 
values of $\beta$. The value of $\lambda$, see Eq. (\ref{2.39}), was fixed 
by calculating the imaginary part of the eigenfrequency from Eq. (\ref{3.3}) 
at $T = 0$ and comparing the result with the experimental value of the 
half-width $\Gamma_{GQR}/2$ of the giant quadrupole resonance in a cold 
nucleus $^{208}Pb$ \cite{sastor77}. 
We find $\lambda/r_0 = 3.011$ for $\beta = 3.4\,{\rm MeV}$ and
$\lambda/r_0 = 3.551$ for $\beta = 9.2\,{\rm MeV}$. The value of $\beta =
1.26\,{\rm MeV}$ is the lower limit for $\beta$ in the sense that, below 
this value, no positive value for $\lambda$ exists such that 
${\rm Im} \,\omega$ equals the experimental value for $\Gamma_{GQR}/2$. 

For each $\beta$ there are three solutions to Eq. (5.3). One of them, 
$i{\rm Im}\,\omega_{in}$, is purely imaginary and lies, for 
$T > T^{(LD)}_{in}$, in the upper half plane of the complex $\omega$ 
(unstable mode). The two other solutions lie symmetric (for small enough 
$T$) with respect to the imaginary axis, $\pm {\rm Re}\,\omega - i 
{\rm Im}\,\omega$, and meet at a temperature $T_{lim}$. For example, we 
find $T_{lim} = 9.5\,{\rm MeV}$ for $\beta = 3.4\,{\rm MeV}$. For larger 
temperatures, $T > T_{lim}$, both solutions are imaginary and 
only the ones nearer to the real axis are shown in Fig. 1. 
Thus, for $T > T_{lim}$ the eigenexcitations exist as overdamped 
modes only. In Fig. 2 we have plotted $T_{lim}$ as function of the 
collisional parameter $\beta$. For $\beta$ larger than about 2.3 MeV, 
the transition between underdamped and overdamped modes occurs in the 
presence of an unstable mode.

In Fig. 3 we have plotted the spectral correlation function 
$\overline{(\alpha_L)_{\omega}^2} $ as obtained from Eq. (\ref{2.43}) for 
the two temperatures $T = 1 \,{\rm MeV}$ and $T = 9 \,{\rm MeV}$. 
The different curves 
show the sensitivity of $\overline{(\alpha_L)_{\omega}^2}$ to the 
parameter $\beta$, Eq. (\ref{2.15}). For low temperature 
we can observe  a well defined maximum which corresponds to the GMR 
excitation (zero-sound regime). An increase of $T$ leads to a shift of 
the maximum of $\overline{(\alpha_L)_{\omega}^2}$ to lower frequencies 
and to an increase in the width. The shape of the curves near 
the zero-sound maximum is a non-Lorentzian one and depends, in particular, 
on the retardation effects in the friction coefficient, Eq. (\ref{2.35}), and, 
consequently, on the parameters $\beta$ and $\zeta$ in the relaxation 
time, Eq. (\ref{2.15}). 
Increasing the temperature  we do not find a first sound peak centered at 
finite frequency for temperatures below $T = T_{in}^{(LD)}$  and for
realistic values of $\beta$, see Fig. 3. The strong increase 
at low frequencies is due to the purely imaginary eigenfrequency.
Thus, there forms no resonance structure of the spectral correlation 
function $\overline{(\alpha_L)_{\omega}^2}$ in this region of $\omega$. 

Moreover, our numerical calculations, with realistic
values of the nuclear parameters, do not show any transition 
from the zero-sound regime at low temperatures to the first-sound regime 
at high temperatures. For the finite Fermi-liquid drop, in the first-sound 
regime the real part of the eigenfrequency is mainly determined by the 
liquid drop stiffness coefficient $C^{(LD)}_L$, i.e., without the additional 
contribution $C^{\prime}_L$ from the Fermi-surface distortion effect. As it 
can be seen from Figs. 1 and 2, even if the limiting temperature $T_{lim}$ 
is well below the temperature $T^{(LD)}_{in}$, the first sound 
regime does not appear because the eigenmotion is overdamped at 
$T_{lim} < T < T^{(LD)}_{in}$.

For a large enough value of $\beta$, i.e. $\beta \geq 2.3 \,{\rm MeV}$,
there is, in principle, a possibility for a resonance structure of 
$\overline{(\alpha_L)_{\omega}^2}$ at temperatures $T > T_{in}^{(LD)}$ 
from the pure Fermi-surface vibration in the momentum space. 
For these values of $\beta$ there exists a temperature region 
$T^{(LD)}_{in} < T < T_{lim}$ where $C_L^{(LD)}(T) \leq 0$ and 
$C_L(\omega_L) > 0$, simultaneously. This implies the existence, in this 
high temperature region, of a particular eigenmode of the Fermi liquid 
drop where the restoring force is exclusively due to the dynamical 
Fermi-surface distortion. Unfortunately, we find this eigenmode  to be
damped too strongly to cause a visible peak in the spectrum of the
fluctuations for values of $\beta < 25 \,{\rm MeV}$. As shown in Fig. 3, 
$\overline{(\alpha_L)_{\omega}^2}$ develops only a weak shoulder 
for the value $\beta = 9.2 \,{\rm MeV}$ at $T = 9 \,{\rm MeV}$. 

\section{Summary and conclusions}

Starting from the collisional Landau-Vlasov kinetic equation
with a random force, we have derived the Langevin equation for the
shape fluctuations in the framework of the nuclear Fermi-liquid drop
model. The main features of these fluctuations are due to the Fermi-surface
distortion effects. We have obtained the random-force correlation
function (\ref{2.37})  for the general case where 
retardation and memory effects in the pressure tensor (\ref{2.11}) and in
the relaxation time (\ref{2.15}) are taken into account. It is important 
to account for these memory effects in order to obtain a correct 
description of the transition  in Fermi-liquid
from the zero-sound regime at low temperature to the 
first-sound regime at high temperatures. We have found, however, that
in a finite nuclear Fermi-liquid the first-sound regime is not
reached. Instead, in the region of reasonable values for our parameters, 
the eigenmotion becomes overdamped at a temperature $T_{lim}$. For 
$T_{lim}$ below the temperature $T^{(LD)}_{in}$ the reappearance of 
underdamped motion is suppressed since the temperature for which 
$C^\prime_L = 0$ is smaller but close to $T_{lim}$. Note also that at 
temperature $T^{(LD)}_{in}$ one of the modes becomes unstable.

Our approach to the shape fluctuations is essentially classical. 
However, due to the quantum version of the fluctuation dissipation
theorem , basic quantum effects are taken into account. 
Thus the correlation functions (\ref{2.21}) and (\ref{2.37}) contain 
contributions from both quantum and thermal fluctuations. 

The effects of the dynamical distortion of the Fermi surface on
the nuclear collective motion lead to peculiarities of the 
random-force correlation function which are absent in a classical
system. The spectral correlation function (\ref{2.37}) is independent
of $\omega$ and corresponds to a white noise in the first-sound
regime at $\omega \tau \rightarrow 0$, whereas in the opposite zero-sound 
regime at $\omega \tau \rightarrow \infty$ it corresponds to a blue
noise (\ref{2.45}). 

The behaviour of the spectral correlation function
$\overline{(\alpha_L)_{\omega}^2}$
at different temperatures reflects the above mentioned 
peculiarities of the random-force correlation function. The
dependence of the shape of the curves $\overline{(\alpha_L)_{\omega}^2}$ 
on the retardation effects ($\omega$-dependence) in the friction coefficient 
(\ref{2.35}) and, consequently, on the parameters $\zeta$ and $\beta$
may serve as means to learn about the possible range of the latter values.  
We find also the existence of an oscillatory mode due to the dynamical 
Fermi-surface distortion in the high temperature region 
$T > T^{(LD)}_{in}$, where the usual liquid-drop
stiffness coefficient $C^{(LD)}_L$ disappears. The spectrum of the 
fluctuations, however, is affected by this mode only for collision 
frequencies which seem too high for stable nuclei.

\section{Acknowledgements}

One of us (V.M.K.) would like to thank the Physics
Department of the TU Munich and Cyclotron Institute at Texas A\&M 
University for the kind hospitality extended to him
during his visits. This work was supported in part by Deutsche
Forschungsgemeinschaft, contract 436 UKR-113/15/0,2463, the US
National Science Foundation under grants \# PHY-9403666 and 
\# PHY 94-13872. We are grateful for this financial support.

\newpage

\begin {center}
{\bf APPENDIX}
\end {center}

In this appendix we are going to determine the correlation function 
of the projected random force $y_{\nu\mu}({\bf r}, t)$ Eq. (\ref{2.9}):
$${1 \over 2}\overline{[y_{\nu\mu}({\bf r}, t); y_{\nu^\prime 
\mu^\prime}({\bf r^\prime}, t^\prime)]}_+ = {1 \over 2}\int 
{d{\bf p}d{\bf p^\prime} \over (2\pi \hbar)^3} p_\nu p_\mu 
p_{\nu^\prime}^\prime p_{\mu^\prime}^\prime {1 \over 2}
\overline{[y_{{\bf p}} ({\bf r}, t); 
y_{{\bf p^\prime}}({\bf r^\prime}, t^\prime)]}_+.
\eqno (A.1)$$
To calculate the correlation function of the random force $y \equiv 
y_{{\bf p}} ({\bf r}, t)$, appearing in Eq. (\ref{2.1}) we follow the 
arguments of Abrikosov and Khalatnikov \cite{ak}. However, we generalize 
the treatment with respect to the following points: we allow for a 
collision term which is {\it (i) non Markovian} and {\it (ii) 
nonlocal in space}:
$$I[\delta n]_{{\bf p},{\bf r};t} = I_{{\bf p},{\bf r};t} * \delta 
\overline{n}_{{\bf p}} ({\bf r},
t)$$ 
$$= \int {dt^\prime d{\bf p^\prime} d{\bf r^\prime} \over 
(2\pi \hbar)^3} {\cal J}({\bf r},{\bf p},t;
{\bf r^\prime},{\bf p^\prime},t^\prime)\delta 
\overline{n}_{{\bf p^\prime}}
({\bf r^\prime}, t^\prime), \eqno (A.2)$$
where
$$\delta \overline{n}_{{\bf p}} ({\bf r}, t) = 
\delta n_{{\bf p}} ({\bf r},t) - 
{\partial n_{p}^0 \over \partial \epsilon_p} \delta \epsilon_{{\bf p}} ,
\eqno (A.3)$$
$n_{p}^0$ is the equilibrium distribution function, $\overline{n}_{p}^0
= 1 - n_{p}^0, \delta \epsilon_{{\bf p}} = 
\sum_{{\bf p^\prime} {\bf r^\prime}} 
f_{{\bf p} {\bf p^\prime}} ({\bf r} - {\bf r^\prime}) 
\delta n_{{\bf p^\prime}}({\bf r^\prime},t)$ 
and $f_{{\bf p} {\bf p^\prime}}({\bf r} - {\bf r^\prime})$ 
is the quasiparticle interaction energy  
\cite{ak}. These generalizations do not change the expression for the rate 
of change of the entropy $\dot{S}$ as compared to \cite{ak}. In the present 
notation we write
$$\dot{S} = - \int {d{\bf r} d{\bf p} d{\bf r^\prime} d{\bf p^\prime} 
\over (2 \pi \hbar)^6}{\delta n_{p} ({\bf r},t) \over 
n_{p}^0 \overline{n}_{p}^0} \Big( h^3 \delta ({\bf p} - {\bf p^\prime}) 
\delta ({\bf r} - {\bf r^\prime}) - {\partial n_{p}^0 \over 
\partial \epsilon_p} f_{{\bf p} {\bf p^\prime}} 
({\bf r} - {\bf r^\prime})\Big)$$ 
$$\times \Big( I[\delta n]_{{\bf p^\prime},
{\bf r^\prime};t^\prime} + y_{{\bf p^\prime}} 
({\bf r^\prime},t^\prime)\Big). \eqno (A.4)$$
In Eq. (A.4) we corrected a typing error in the second term of the right 
hand side of Eq. (11.9) of Ref. \cite{ak}.

To apply the general theory of fluctuations \cite{ll} we need to write
$\dot{S}$ in the form
$$\dot{S} = - \sum_{i} X_i (t) \dot{x}_i (t).  \eqno (A.5)$$
In the present application the index $i$ stands for the pair of
vectors ${\bf p},{\bf r}$ and the sum has to be replaced by the 
integral over phase space. Defining
$$\dot{x}_i (t) = I[\delta n]_{{\bf r}_i,{\bf p}_i;t} + 
y_{{\bf p}_i}({\bf r}_i,t), 
\eqno (A.6)$$
the expression (A.3) for $\dot{S}$ takes the form (A.5) if we set 
the generalized force $X_i(t) \equiv X ({\bf r}_i,{\bf p}_i;t)$  as
$$X ({\bf r}_i,{\bf p}_i;t) = {1 \over n_{p_i}^0 \overline{n}_{p_i}^0} 
[\delta n_{{\bf p}_i}({\bf r}_i,t) - {\partial n_{p_i}^0 \over \partial 
\epsilon_{p_i}}\delta \epsilon_{{\bf p}_i} ({\bf r}_i,t)] = 
{\delta \overline{n}_{{\bf p}_i} ({\bf r}_i,t) \over 
n_{p_i}^0 \overline{n}_{p_i}^0}. \eqno (A.7)$$
In the next step, we have to express the first term on the right hand
side  of Eq. (A.6) in terms of the $X({\bf r},{\bf p}; t)$, Eq. (A.7).
With the help of Eq. (A.2) we find:
$$I[\delta n]_{{\bf r},{\bf p};t} = I_{{\bf r},{\bf p};t} * 
\Big( n_{p}^0 \overline{n}_{p}^0 
X({\bf r},{\bf p}; t)\Big). \eqno (A.8)$$
Comparing the general form of $\dot{x}_i (t)$, generalized to non
Markovian processes \cite{fick}
$$\dot{x}_i (t) = -\sum_{j,t^\prime} \gamma_{ij} (t, t^\prime) 
X_j (t^\prime) + y_i (t)   \eqno (A.9)$$
with Eqs. (A.6) and (A.8) we can easily determine the coefficients
$\gamma_{ij} (t,t^\prime)$
$$\gamma ({\bf r},{\bf p},t ; {\bf r^\prime},{\bf p^\prime},t^\prime) =
{\cal J} ({\bf r},{\bf p},t ; {\bf r^\prime},{\bf p^\prime},t^\prime)
n_{p^\prime}^0 \overline{n}_{p^\prime}^0. \eqno (A.10)$$
If we assume that $\cal{J}$ depends on the difference $t - t^\prime$
only, we obtain for the correlation function of the random force
$${1 \over 2}\overline{[y_{{\bf p}} ({\bf r},t) ; 
y_{{\bf p^\prime}} ({\bf r^\prime},t)]_+} =
\int {d\omega \over 2\pi} e^{-i\omega (t-t^\prime)} E(\omega,T) $$
$$\times [{\cal J} ({\bf r},{\bf p};{\bf r^\prime},{\bf p^\prime};
\omega) {\partial n_{p^\prime}^0 \over \partial \epsilon_{p^\prime}} + 
{\cal J} ({\bf r^\prime},{\bf p^\prime};{\bf r},{\bf p};-\omega) 
{\partial n_{p}^0 \over \partial \epsilon_p}], \eqno (A.11) $$
where $E(\omega, T)$ is determined by Eq. (\ref{2.22}). This expression
is valid for a quite general collision term. In the present paper
we do not need the full correlation function, but only its 
projection (A.1). For the latter we find
$${1 \over 2}\overline{[y_{\nu \mu}({\bf r},t); y_{\nu^\prime 
\mu^\prime}({\bf r^\prime},t^\prime)]}_+ = {1 \over m^2}\int {d\omega 
\over 2\pi} e^{-i\omega(t-t^\prime)} E(\omega,T) \int 
{d{\bf p} d{\bf p^\prime} \over (2\pi\hbar)^6}p_\nu p_\mu $$
$$\times[{\cal J}({\bf r},{\bf p}; 
{\bf r^\prime},{\bf p^\prime}; \omega)
{\partial n_{p^\prime}^0 \over \partial \epsilon_{p^\prime}} + 
{\cal J}({\bf r^\prime},{\bf p^\prime}; {\bf r},{\bf p}; -\omega)
{\partial n_{p}^0 \over \partial \epsilon_p}]p_{\nu^\prime}^\prime 
p_{\mu^\prime}^\prime. \eqno (A.12)$$

The collision integral used in the text is assumed to have the 
properties (\ref{2.14}) with (\ref{2.8}). From the latter we find the kernel
$\cal J$ in (A.2) to fulfil the relation:
$$\int {d{\bf p} \over (2\pi\hbar)^3} \int {d{\bf p^\prime} \over 
(2\pi\hbar)^3} p_{\nu}^\prime p_{\mu}^\prime
{\cal J}({\bf r^\prime},{\bf p^\prime}; {\bf r},
{\bf p}; \omega) {\partial n_{p}^0 \over \partial \epsilon_p}
(p_{\nu^\prime} p_{\mu^\prime} - {1 \over 3}\delta_{\nu^\prime 
\mu^\prime}p^2) $$
$$= {1 \over \tau (\omega)} \int {d{\bf p} \over 
(2\pi\hbar)^3}p_\nu p_\mu (-{\partial n_{p}^0 \over \partial 
\epsilon_{p}})(p_{\nu^\prime}p_{\mu^\prime} - {1 \over 3}
\delta_{\nu^\prime \mu^\prime}p^2)\delta ({\bf r} - {\bf r^\prime}),
\eqno (A.13)$$
whereas
$$\int {d{\bf p} \over (2\pi\hbar)^3}{\cal J}({\bf r^\prime}, 
{\bf p^\prime};{\bf r},{\bf p}; \omega) {\partial n_{p}^0 \over \partial 
\epsilon_p} = 0  $$
due to conservation of particle number in collisions. With the help of
$$\int {d{\bf p} \over (2\pi\hbar)^3}p_\nu p_\mu (-{\partial n_{p}^0 
\over \partial \epsilon_p})(p_{\nu}^\prime p_{\mu}^\prime - {1 \over 3} 
\delta_{\nu^\prime \mu^\prime} p^2) $$
$$= (\delta_{\nu \nu^\prime}
\delta_{\mu \mu^\prime} + \delta_{\nu \mu^\prime} \delta_{\nu^\prime 
\mu} - {2 \over 3}\delta_{\nu \mu}\delta_{\nu^\prime \mu^\prime})
\int {d{\bf p} \over (2\pi\hbar)^3} {p^4 \over 15}(-{\partial n_{p}^0 
\over \delta \epsilon_p})  \eqno (A.14)$$
one finds
$${1 \over 2}\overline{[y_{\nu\mu}({\bf r},t); y_{\nu^\prime \mu^\prime}
({\bf r^\prime},t^\prime)]}_+ = \int {d\omega \over 2\pi}
e^{-i\omega(t-t^\prime)} E(\omega, T)[{1 \over \tau (\omega)} +
{1 \over \tau (-\omega)}]$$
$$\times \int {d{\bf p} \over (2\pi\hbar)^3} 
{p^4 \over 15m^2}(-{\partial n_{p}^0 \over \partial \epsilon_p})
(\delta_{\nu \nu^\prime}              
\delta_{\mu \mu^\prime} + \delta_{\nu \mu^\prime} \delta_{\nu^\prime 
\mu} - {2 \over 3}\delta_{\nu \mu}\delta_{\nu^\prime \mu^\prime})
\delta ({\bf r} - {\bf r^\prime}). \eqno (A.15)$$
Here
$$\int {d{\bf p} \over (2\pi\hbar)^3} 
{p^4 \over 15m^2}(-{\partial n_{p}^0 \over \partial \epsilon_p}) =
P_{eq}. \eqno (A.16)$$
In Fourier space the correlation function is given from (A.15) as
$${1 \over 2}\overline{[y_{\nu\mu}({\bf r},\omega); 
y_{\nu^\prime \mu^\prime}({\bf r^\prime},\omega^\prime)]}_+$$
$$= 4\pi \delta (\omega + \omega^\prime) \delta ({\bf r} - {\bf r^
\prime}) E(\omega,T){P_{eq} \over \tau (\omega)} (\delta_{\nu \nu^
\prime}\delta_{\mu \mu^\prime} + \delta_{\nu \mu^\prime} \delta_{\nu^
\prime \mu} - {2 \over 3}\delta_{\nu \mu}\delta_{\nu^\prime 
\mu^\prime})
\eqno (A.17)$$
and the one for the random pressure tensor (\ref{2.20}) reads
$${1 \over 2}\overline{[s_{\nu\mu}({\bf r},\omega); 
s_{\nu^\prime \mu^\prime}({\bf r^\prime},\omega^\prime)]}_+ $$
$$= 4\pi \delta (\omega + \omega^\prime) \delta ({\bf r} - 
{\bf r^\prime}) E(\omega,T){\tau (\omega) \over
{1 + (\omega \tau)^2}}{P_{eq} \over m^2}(\delta_{\nu \nu^\prime}
\delta_{\mu \mu^\prime} + \delta_{\nu \mu^\prime} \delta_{\nu^\prime\mu} 
- {2 \over 3}\delta_{\nu \mu}\delta_{\nu^\prime \mu^\prime}).
\eqno (A.18)$$
With the definition of the viscosity coefficient (\ref{2.19}) we obtain 
formula (\ref{2.21}).

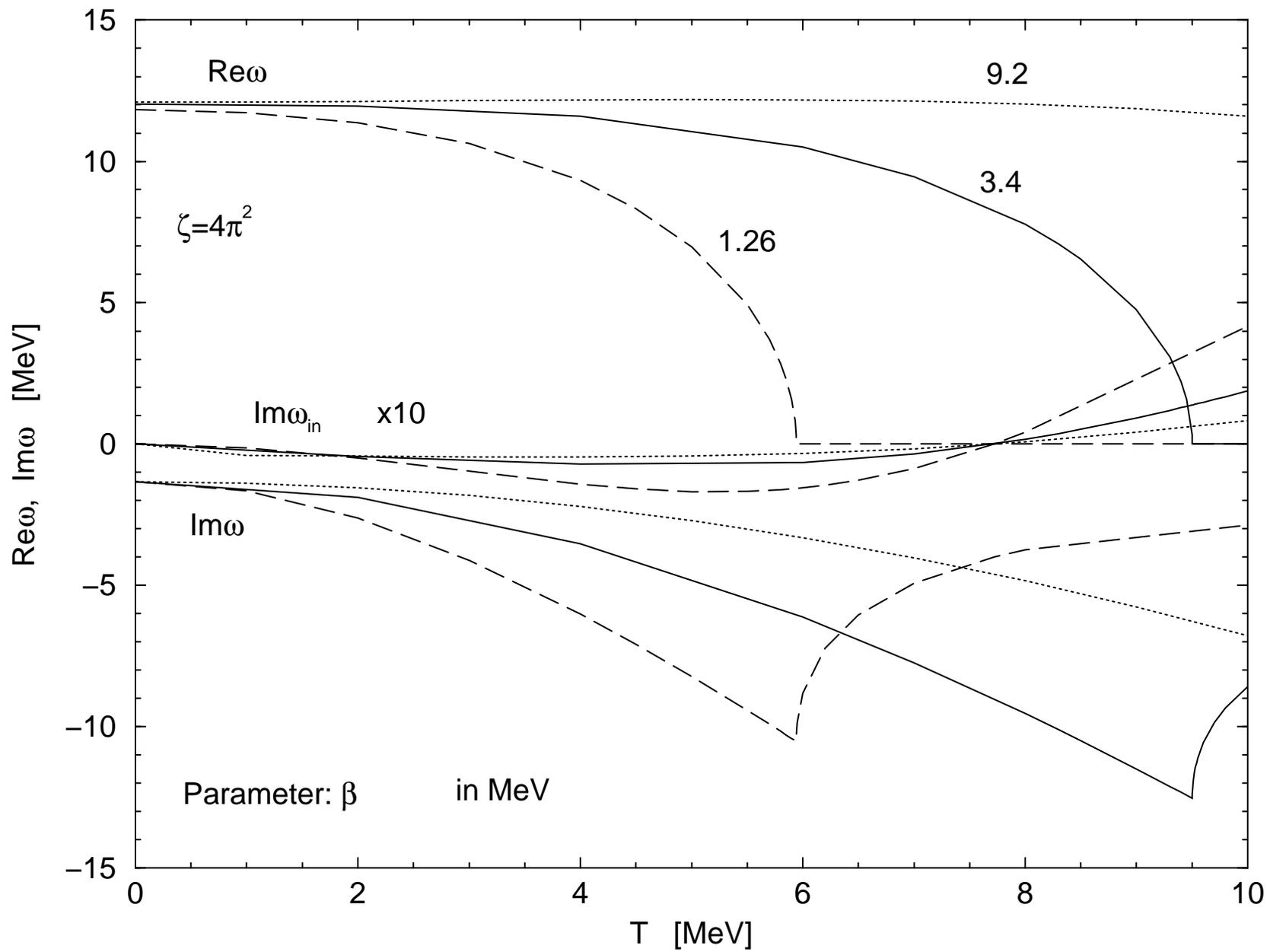
\begin{figure}
\caption{The dependence of the   real, ${\rm Re}\,\omega$, and
imaginary, ${\rm Im}\,\omega$ parts of the eigenfrequency $\omega$
from Eq. (5.3) on the temperature $T$ for a nucleus $A = 208$
for the quadrupole vibrations $L = 2$; the relaxation time parameters
$\beta = 1.26 \,{\rm MeV}$ (dashed lines), \,$3.4 \,{\rm MeV}$ (solid
lines) and $\beta = 9.2 \,{\rm MeV}$ (dotted lines) are shown as 
a label to the curves; the parameter $\zeta$ is chosen as $\zeta = 
4\pi^2$.}
\end{figure}

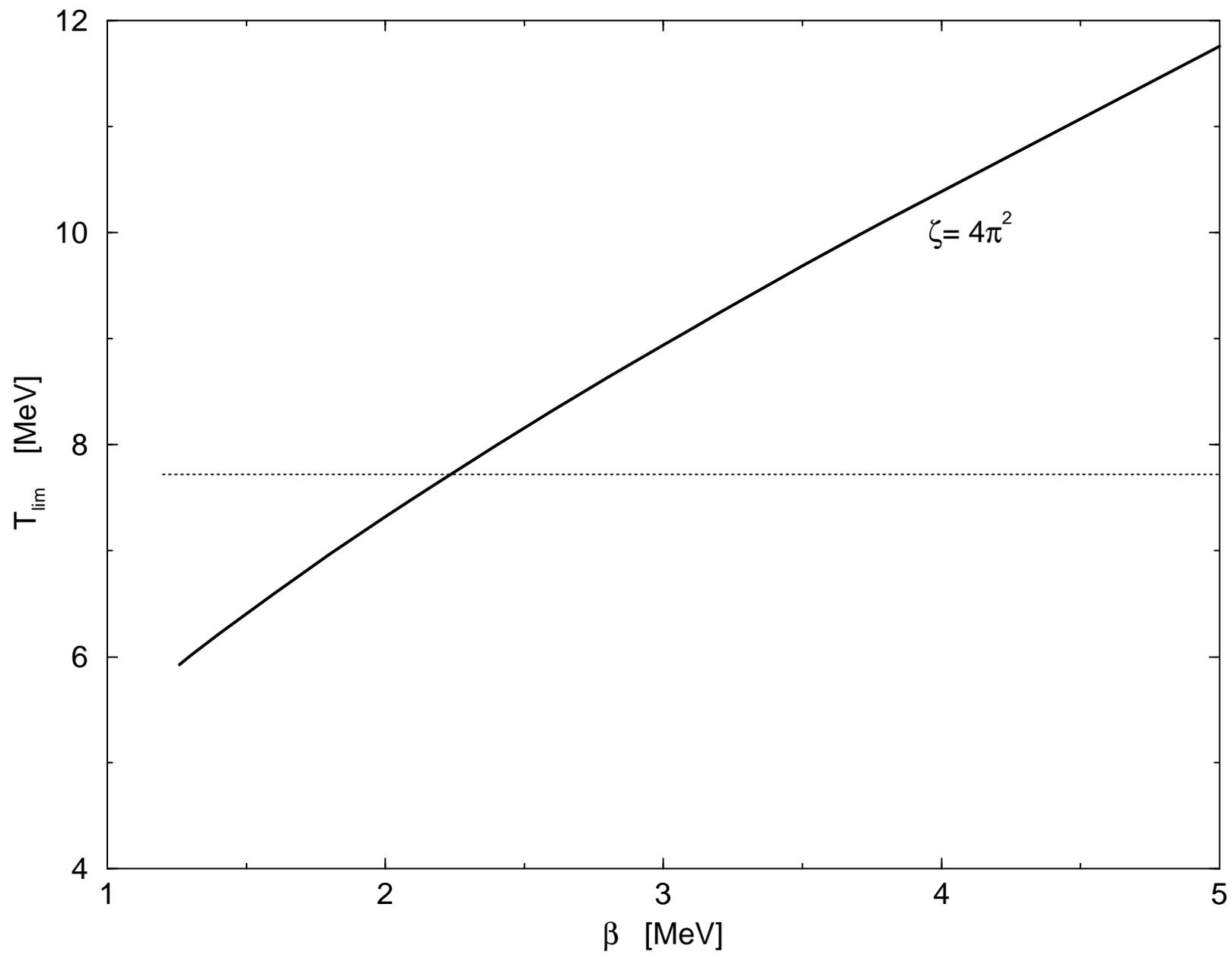
\begin{figure}
\caption{The limiting temperature $T_{lim}$ as a function 
of the relaxation parameter $\beta$; using $\zeta = 4\pi^2$. 
The doted line is the critical temperature
$T^{(LD)}_{in}$ where the liquid drop stiffness coefficient
$C^{(LD)}_L$ vanishes.}
\end{figure}

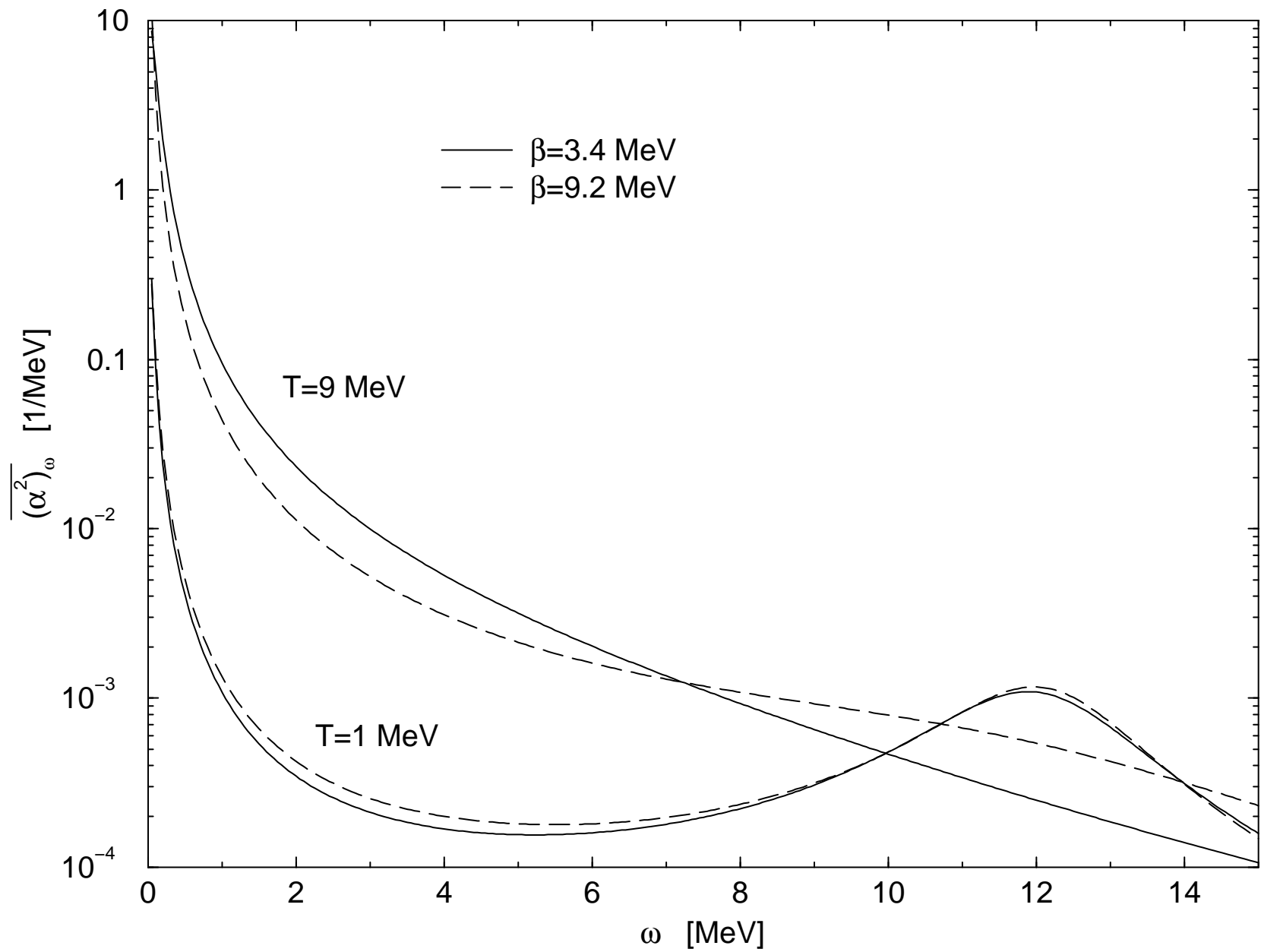
\begin{figure}
\caption{The spectral correlation function
$\overline{(\alpha_L)_{\omega}^2}$ for two temperatures: $T=1 \,{\rm MeV}$
and $T=9\, {\rm MeV}$. The calculations were performed
for $\zeta = 4\pi^2$ and the values $\beta = 3.4\, {\rm MeV}$ and 
$\beta = 9.2\, {\rm MeV}$.}
\end{figure}

\end{document}